# Synthesizing gas-filled fiber Raman lines enables access to the molecular fingerprint region

Yazhou Wang,[1,*] Lujun Hong,[1,†] Cuiling Zhang,[1] Joseph Wahlen,[2] J. E. Antonio-Lopez,[2] Manoj K. Dasa,[3] Abubakar I. Adamu,[1,4] Rodrigo Amezcua-Correa,[2] and Christos Markos[1,5,*]

[1]DTU Electro, Technical University of Denmark, 2800 Kgs. Lyngby, Denmark

[2]CREOL, The College of Optics and Photonics, University of Central Florida, Orlando, Florida 32816, USA

[3]NKT Photonics A/S, Blokken 84, Birkerød DK-3460, Denmark

[4]Microsoft Azure, Unit 7, The Quadrangle, Romsey, S051 9DL, UK

[5]NORBLIS ApS, Virumgade 35D, 2830 Virum, Denmark

*Correspondence to yazwang@dtu.dk & chmar@dtu.dk

[†]Now affiliated with: Institute of Space Science and Technology, Nanchang University, Nanchang 330031, China

**Abstract:** The synthesis of multiple narrow optical spectral lines, precisely and independently tuned across the near- to mid-infrared (IR) region, is a pivotal research area that enables selective and real-time detection of trace gas species within complex gas mixtures. However, existing methods for developing such light sources suffer from limited flexibility and very low pulse energy, particularly in the mid-IR domain. Here, we introduce a new concept based on the gas-filled anti-resonant hollow-core fiber (ARHCF) technology that enables the synthesis of multiple independently tunable spectral lines with high pulse energy of >1 μJ and a few nanoseconds pulse width in the near- and mid-IR region. The number and wavelengths of the generated spectral lines can be dynamically reconfigured. A proof-of-concept laser beam synthesized of two narrow spectral lines at 3.99 μm and 4.25 μm wavelengths is demonstrated and combined with photoacoustic (PA) modality for real-time $SO_2$ and $CO_2$ detection. The proposed concept also constitutes a promising way for IR multispectral microscopic imaging.

# Introduction

The IR region, known as the molecular fingerprint region, plays an essential role in a wide arc of applications particularly for trace gas analysis. While diverse IR light sources have been developed [1–5], currently the research community focuses on developing next-generation multi-analyte gas sensors with high sensitivity, high selectivity, and fast response time. A promising solution to meet these requirements, is the synthesis of multiple narrow spectral lines, where each distinct laser line targets a specific gas absorption without interference from other gas species. However, developing such a light source remains a major scientific challenge. The conventional laser diodes at different wavelengths can be synthesized into a single-mode silica fiber through wavelength division multiplexer (WDM), but this approach is limited to the near-IR region (< 2.4 µm) [6]. To move into the mid-IR region, the only approach is of synthesizing multiple quantum cascade lasers (QCLs) with different emission wavelengths [7–12]. Specifically, a mid-IR photonic integrated circuit has been recently proposed to synthesize spectral lines from multiple QCL sources [9,13]. However, QCL technology suffers from low (peak) power of just a few watts [14–17], thus is fundamentally limited towards the generation of high-energy mid-IR pulses [18]. Additionally, this technology cannot dynamically tailor the number and wavelengths of synthesized spectral lines.

The recent advent of gas-filled ARHCF Raman laser technology [19–22] is a potential alternative for the synthesis of IR spectral lines with high pulse energy. The long Raman Stokes coefficient of active gases [23] combined with the broadband and low-loss light propagation of ARHCF [24], enable the efficient Raman Stokes wavelength conversion from the near-IR to the mid-IR region. As a result, several near- and mid-IR gas-filled fiber Raman lasers have been reported with microjoule-level high energy nanosecond pulses up to 4.4 µm wavelength [20,21,25–29]. Meanwhile, the Raman gain width can down to a few GHz even hundreds MHz levels by using specific gases at appropriate pressure [30–32], thus allowing the generation of narrow Raman spectral line(s) [26,33,34]. However, the existing gas-based Raman lasers have only been developed based on single-wavelength pump lasers with pre-defined static Raman line(s) and thus independent wavelength tuning is inherently not possible.

In this work, we propose a novel concept of employing a multi-seed laser as a pump coupled into a cascaded configuration of gas-filled ARHCFs, to enable the generation of multiple reconfigurable narrow Raman spectral lines spanning from near- to mid-IR region. The number and wavelengths of Raman lines can be independently and precisely controlled/tuned, to target the absorption lines of selected gas species.

**Concept for the synthesis of multiple spectral lines**

Figure 1(a) presents the proposed concept of synthesizing multiple spectral lines. The center part is a cascaded silica ARHCF configuration pumped by a near-IR laser with multiple synthesized reconfigurable spectral lines lying within the gain range of Ytterbium(Yb)-doped fiber amplifier (typically 1015 to 1115 nm [35]). The wavelengths of these synthesized pump lines are nonlinearly converted based on the rotational/vibrational stimulated Raman scattering (SRS) effect of gases filled in the 1st and 2nd stage ARHCFs. As a result, the synthesized Raman Stokes lines have diverse wavelengths from the near- to mid-IR region. The wavelength of each Raman Stokes line is determined by the wavelength of its corresponding pump line through $\frac{1}{\lambda_R} = \frac{1}{\lambda_P} - \Omega$, where $\lambda_p$ and $\lambda_R$ are the optical wavelengths of the pump and Raman Stokes, respectively, $\Omega$ is the rotational/vibrational Raman shift coefficient of gas.

The key of this concept is the broad IR wavelength range of the Raman Stokes lines enabling the access to the absorption bands of different gases. The wavelength ranges of Raman Stokes and pump laser can be expressed as:

$$\Delta\lambda_R = \lambda_{R\_max} - \lambda_{R\_min} = \frac{\Delta\lambda_P}{(1-\Omega\lambda_{P1})(1-\Omega\lambda_{P2})} \tag{1}$$

where $\Delta\lambda_P = \lambda_{P\_max} - \lambda_{P\_min}$ is the pump wavelength range, $\Delta\lambda_R = \lambda_{R\_max} - \lambda_{R\_min}$ is the corresponding Raman Stokes wavelength range, $\Omega$ is the Raman Stokes shift coefficient. With a large $\Omega$ value, the denominator in equation (1) is far less than 1, thus making $\Delta\lambda_R$ much greater than $\Delta\lambda_P$ (see examples in Supplementary Table 1). In the case of cascading two ARHCFs, $\Omega = \Omega_1 + \Omega_2$, where $\Omega_1$ and $\Omega_2$ are the Raman Stokes shift coefficients of gases filled in the 1$^{st}$ and 2$^{nd}$ stage ARHCF, respectively.

While any pump wavelength can be selected, we specifically select two pump lines at 1044 and 1060 nm, to demonstrate the concept behind Equation (1). These two lines are used to pump two cascaded silica ARHCFs (see Supplementary Note 2). Figure 1b shows the diversity of the generated and synthesized Raman lines over a broad IR range from 1.2 to 4.3 μm, achieved by filling 30 bar $H_2$ into the 2nd stage ARHCF while the 1st stage ARHCF is filled with different gases including $CH_4$, $CO_2$, and $N_2$, respectively. These results show that each pair of Raman Stokes lines has a larger spacing than the dual pump lines (see details in Supplementary Note 1). For example, synthesized Raman laser lines at 3.99 and 4.25 μm are generated in the case of filling $CH_4$ into the 1st stage ARHCF. The spacing between these two Raman lines is approximately 15 times larger than that of the dual pump lines, therefore the Equation (1) is validated. Furthermore, the dedicated selection of the pump wavelengths makes the synthesized dual Raman lines (at 3.99 and 4.25 μm) exactly overlap with the absorption spectra of $SO_2$ and $CO_2$, respectively, to enable selective and real-time detection of both gases.

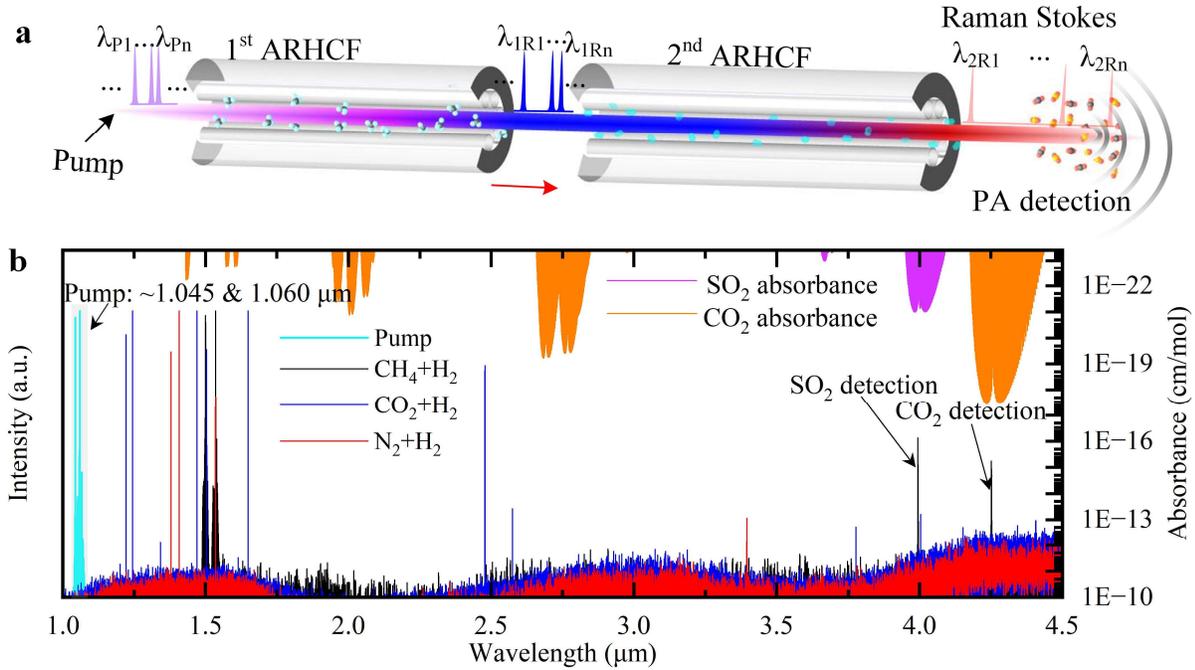

**Fig. 1 | Concept illustration. a** Concept for the synthesis of multiple IR spectral lines and its application on PA detection of multiple gases. $\lambda_{P1}...\lambda_{Pn}$ refer to the wavelengths of pump lines, $\lambda_{1R1}...\lambda_{1Rn}$ refer to the wavelengths of Raman laser lines output from the 1$^{st}$ gas-filled ARHCF, $\lambda_{2R1}...\lambda_{2Rn}$ refer to the wavelengths of Raman laser lines output from the 2$^{nd}$ gas-filled ARHCF. **b** Wavelength conversion examples through filling different gas ($CH_4$, $CO_2$ and $N_2$) into the 1$^{st}$ stage ARHCF. The 2$^{nd}$ stage ARHCF is constantly filled with $H_2$ at 30 bar. Left axis shows measured optical spectra including pump lines, and Raman lines output from the $H_2$-filled 2$^{nd}$ stage ARHCF. Right axis shows the absorbance spectra of $CO_2$ and $SO_2$ obtained from the high-resolution transmission molecular absorption database (HITRAN). Details regarding the measured spectra in **b** is explained in Supplementary Note 1.

## Design, characterization, and photoacoustic gas-detection application of the proposed concept

This section consists of A): the synthesis of multiple spectral lines using the proposed concept, and B): the application of the proposed laser source on PA detection of multiple gases. Figure 2a shows the configuration of the entire system.

### A. The synthesis of multiple spectral lines using the proposed concept

This part consists of the pump fiber laser and the cascaded ARHCF Raman laser.

A.1 Design and characterization of the pump laser

The pump laser adopts an all-fiber polarization-maintained (PM) structure. The seed laser is composed of two linearly polarized distributed feedback narrow-linewidth (few GHz) laser diodes with different wavelengths at 1044 and 1060 nm which are well within the gain range of Yb-doped fiber. These two seeds have a single-mode fiber (PM980) output, and they are synthesized into a single-mode PM fiber through WDM, to seed the subsequent PM Yb-doped fiber amplifiers, as shown in the top of Fig. 2a. The number and wavelength of the laser diode seeds can be dynamically added/removed through fiber splicing. All laser diode seeds are directly modulated, to emit 3.7 ns pulses. Each seed is equipped with a thermoelectric cooler for thermally and independently tuning its wavelength over ~1 nm in the temperature range of 20 to 35 °C. The amplification module is composed of two pre-amplifiers and a power amplifier using 915 nm laser diodes as pumps (see Methods section). The Yb-doped fiber in the power amplifier has a large core diameter of 25 μm, to suppress the stimulated Brillouin Scattering (SBS) effect. To avoid the gain competition during the amplification, a trigger-delay system is designed to make an alternate output between different spectral lines in the time domain, as shown by an example at the top of Fig. 2b, where each line operates in burst mode, composed of a certain number of internal pulses with a certain repetition rate which can be set through the control software. The interval between adjacent bursts is set to 4 ms, which is sufficiently longer than the upper-level lifetime of the $Yb^{3+}$ ion (in the order of 1-2 ms) and thus avoids the gain competition of different spectral lines. Such a short time interval guarantees a fast switching of the spectral lines of the pump and thus the subsequent Raman lines (see an example at the bottom of Fig. 2b), thereby constituting the proposed concept of "real-time detection of multiple gases".

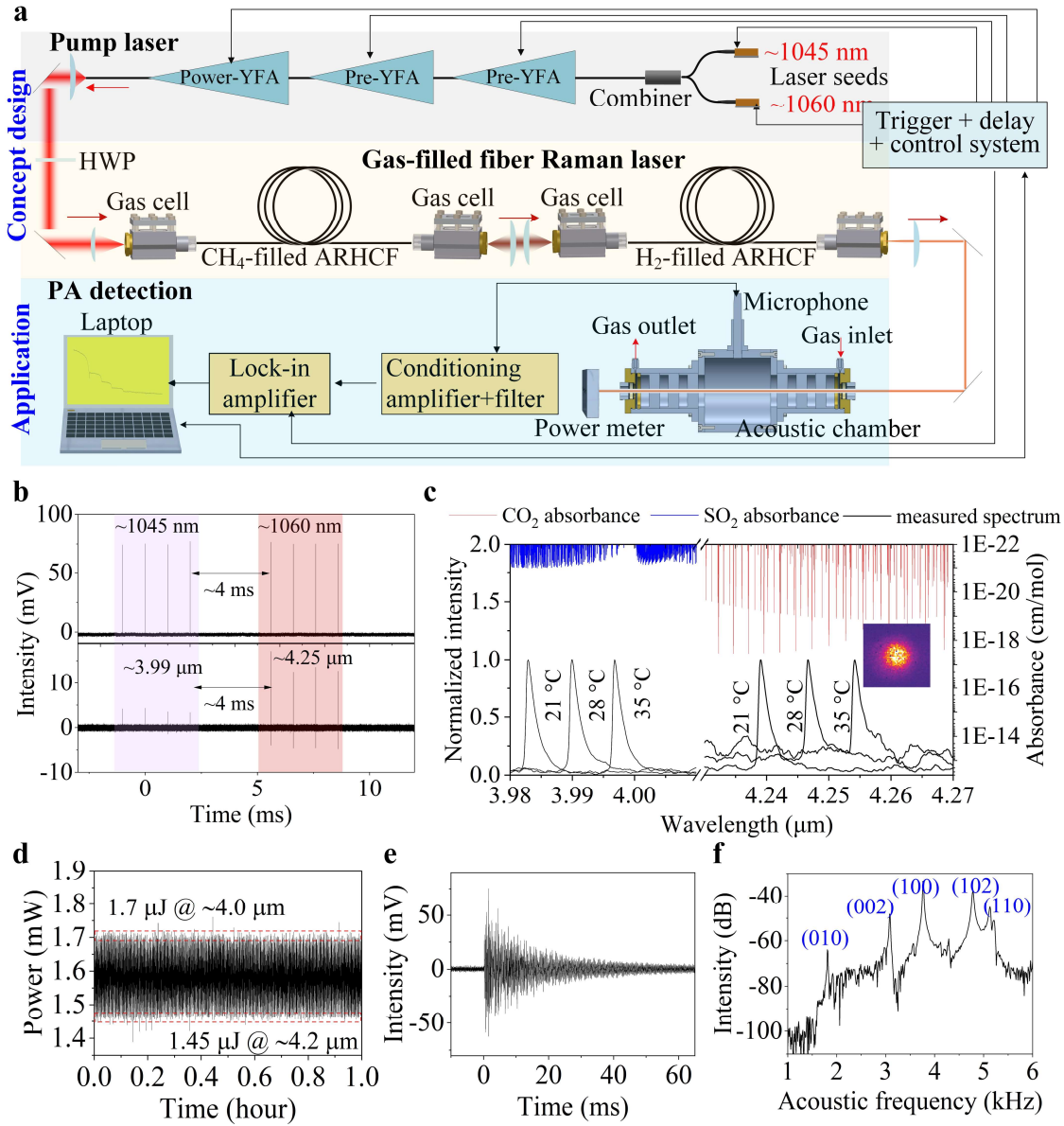

**Fig. 2 | Experiment setup and main characterization results. a** Experiment setup consisting of the pump laser, cascaded ARHCFs filled with $CH_4$ and $H_2$, respectively, PA detection, and data acquisition. HWP: half-wave plate. YFA: Yb-doped fiber amplifier. **b** Top: pump pulse bursts alternately output between 1044 and 1060 nm; bottom: corresponding Raman pulse bursts alternately output between 3.99 and 4.25 μm from the 2$^{nd}$ stage $H_2$-filled ARHCF. **c** Independent and precise wavelength tuning of the mid-IR Raman Stokes lines by thermally tuning the wavelengths of their corresponding laser diode seeds at 1 μm region. The right axis shows the absorbance spectra of $CO_2$ and $SO_2$ obtained from HITRAN. Inset of **c** is the measured beam profile of 4.25 μm Raman laser using a mid-IR camera. **d** Average power monitoring of the Raman laser alternately output between 3.99 and 4.25 μm. **e** PA pulse excited at 1% $CO_2$ concentration with 4.25 μm Raman laser at 10 Hz repetition rate. **f** Fourier transformation spectrum of **e**. (n, m, $n_z$) in **f** refers to different radial, azimuthal, and longitudinal modes of the resonant acoustic chamber [38].

The pulse pump regime [36] is adopted for the Yb-doped fiber amplifier module by electrically modulating the 915 nm laser diode pumps, and the intensity and width of the pump pulse can be independently adjusted for different laser lines to be amplified. This design brings two key advantages. i) It suppresses the unwanted amplified spontaneous emission (ASE) without using bandpass filters in the amplification stages, therefore enabling the "reconfigurable property" of synthesizing any number and wavelength of laser diode seeds within the Yb gain range. ii) Given the varied gain coefficient of the Yb-doped fiber amplifier with respect to wavelength, for different laser lines this design allows for using different pump parameters to maximize their output pulse energies respectively. In this design, the maximum internal repetition rate of the laser burst is limited to ~1 kHz due to the ~2 ms upper-level lifetime of the $Yb^{3+}$ ion.

After amplification, the maximum pulse energy at 1044 nm and 1060 nm are measured to be ~75 and ~98 µJ, respectively, using a pyroelectric energy meter (PE9-ES-C, Ophir Optronics), while the linewidths are less than 0.2 nm which is important for suppressing the dispersion walk-off effect and improving the efficiency of Raman Stokes generation [37]. The polarization extinction ratio of the amplified spectral lines is measured to be ~20 dB which slightly varies as a function of the temperature of corresponding laser diode seed. The detailed characteristics of the pump laser are provided in Supplementary Note 3.

A.2 Design and characterization of the cascaded gas-filled ARHCF Raman laser

As shown in Fig. 2a, the pump laser is coupled into the 1st stage 7-m long $CH_4$-filled ARHCF through a pair of C-coated plano-convex lenses, to enable the synthesis of a dual Raman laser at 1.50 and 1.53 µm through the 1st order vibrational Raman Stokes generation of $CH_4$. The $CH_4$ pressure is set to 2.5 bar, to achieve a relatively narrow linewidth of ~0.5 nm with a high pulse energy. The average pulse energies of the 1st order vibrational Raman Stokes lines are respectively measured to be 22 µJ at 1.50 µm and 28 µJ at 1.53 µm and their pulse widths are measured to be 3.3 ns due to the short dephasing time of $CH_4$. Details regarding the characterization of the $CH_4$-filled ARHCF Raman laser are included in Supplementary Note 4.

The two Raman Stokes lines output from the 1st stage $CH_4$-filled ARHCF is then coupled into the 2nd stage $H_2$-filled ARHCF with 5 m length. The two Raman lines at 3.99 and 4.25 µm are generated and synthesized through the 1st order vibrational Raman Stokes generation in $H_2$ gas. Due to the pump laser design, the two synchronized Raman laser lines alternately occur in the time domain (see the example at the bottom of Fig. 2b). Figure 2c presents the measured optical spectra. Their linewidths are measured to be ~1.3 nm using an optical spectrometer (Spectro 320, Instrument Systems) with a resolution of 0.3 nm. By thermally tuning their corresponding seed line, the wavelength of each Raman line can be precisely and independently tuned over ~15 nm, to target the absorption lines of $SO_2$ and $CO_2$. Here, the temperatures for the 1044 and 1060 nm seeds are set to 22 °C and 24 °C, respectively, to maximize the absorption coefficient (see Fig. 2c) and thus the excited acoustic signal during the subsequent PA gas detection process. Under this condition, the maximum pulse energies are measured to be 1.7 µJ at 3.99 µm and 2.7 µJ at 4.25 µm, respectively, at 30 bar $H_2$ pressure. They correspond to a quantum efficiency of 6.0% and 7.6% in terms of the pump energies of 75 µJ at 1044 nm and 98 µJ at 1060 nm. Figure 2d presents the power monitoring result using a thermal power meter. In this measurement, the burst of each Raman line in the time domain is composed of 1000 pulses with a 1 kHz internal repetition rate. The periodic switching of the measured power indicates the alternate output between the synthesized mid-IR Raman lines. Note that the result in Fig. 2d shows a 1.5 µJ pulse energy at the 4.25 µm line which is less than the aforementioned 2.7 µJ because of the laser beam attenuation caused by $CO_2$ absorption in the ambient air (see Supplementary Fig. S5c). Details regarding the $2^{nd}$ stage $H_2$-filled fiber Raman laser are referred to Supplementary Note 5.

**B. Photoacoustic detection of multi-gases**

This part consists of the configuration of the PA detection setup, and real-time detection of $CO_2$ and $SO_2$.

B.1 Configuration of the PA detection setup

The Raman laser beam consisting of synthesized 3.99 and 4.25 µm Raman spectral lines is collimated by a $CaF_2$ lens with a 3 cm focal length and then coupled into a resonant acoustic chamber for $SO_2$ and $CO_2$ detection. The acoustic chamber adopts the design in Ref. 38 & 39 (see the cross-section of the acoustic chamber at the bottom of Fig. 2a), which has widely used nanosecond pulsed laser as a light source [18,38,40]. Details regarding the PA detection setup and gas sample preparation are described in Methods. Figure 2e shows a typical acoustic pulse excited by the 4.25 µm gas-filled fiber laser operating at a 10 Hz repetition rate, where 1% $CO_2$ is used by diluting a pure $CO_2$ with pure Ar. It shows a sudden onset of an oscillation followed by a slow decay in amplitude, a typical characteristic of an acoustic wave excited from a nanosecond laser pulse [38]. The pulse width of the acoustic wave is measured to be ~7 ms. The corresponding Fourier transformation spectrum is shown in Fig. 2f. The sharp peaks are the resonant modes of the acoustic chamber, indicating the broad frequency range of the acoustic pulse excited by the nanosecond Raman laser pulse, which induces the advantage of robust PA performance against the variation of the ambient temperature variation [18].

B.2 Real-time PA detection of $SO_2$ and $CO_2$

$SO_2$ and $CO_2$ mixture diluted by pure Ar flows through the acoustic chamber for validating the proposed detection approach. Their concentrations are independently varied by mass flow controllers (MFCs) (see details in Methods). Every pulse burst of the mid-IR Raman laser lines is set to be composed of 1000 periodic Raman pulses, leading to a periodic excitation of the PA signal which is detected in the frequency domain using a digital lock-in amplifier (MFLI 500 kHz, Zurich Instruments). First, the internal repetition rate of the laser burst is set to 942 Hz, to overlap its 4$^{th}$-order harmonic frequency component of the acoustic pulse sequence with the first radial mode (100) of the acoustic chamber at 3.77 kHz, as shown in Fig. 2f. In this case, the PA detection combines both the modulation and pulse regimes, leading to an enhanced sensitivity [41].

Figure 3a presents the raw acoustic data of the detection of $SO_2$ and $CO_2$ obtained from the lock-in amplifier, where both of their concentrations vary as a function of time. The zoom-in in the inset shows that the signal switches in ~2 s time interval because of the alternative output of Raman laser lines at 3.99 and 4.25 μm. This ~2 s time interval is dependent on the response speed of the lock-in amplifier determined by the time constant, which is set to 30 ms. Using a shorter time constant, a quicker response speed can be achieved but at the price of increasing the fluctuations of the measured acoustic signal and thus compromising the detection limit.

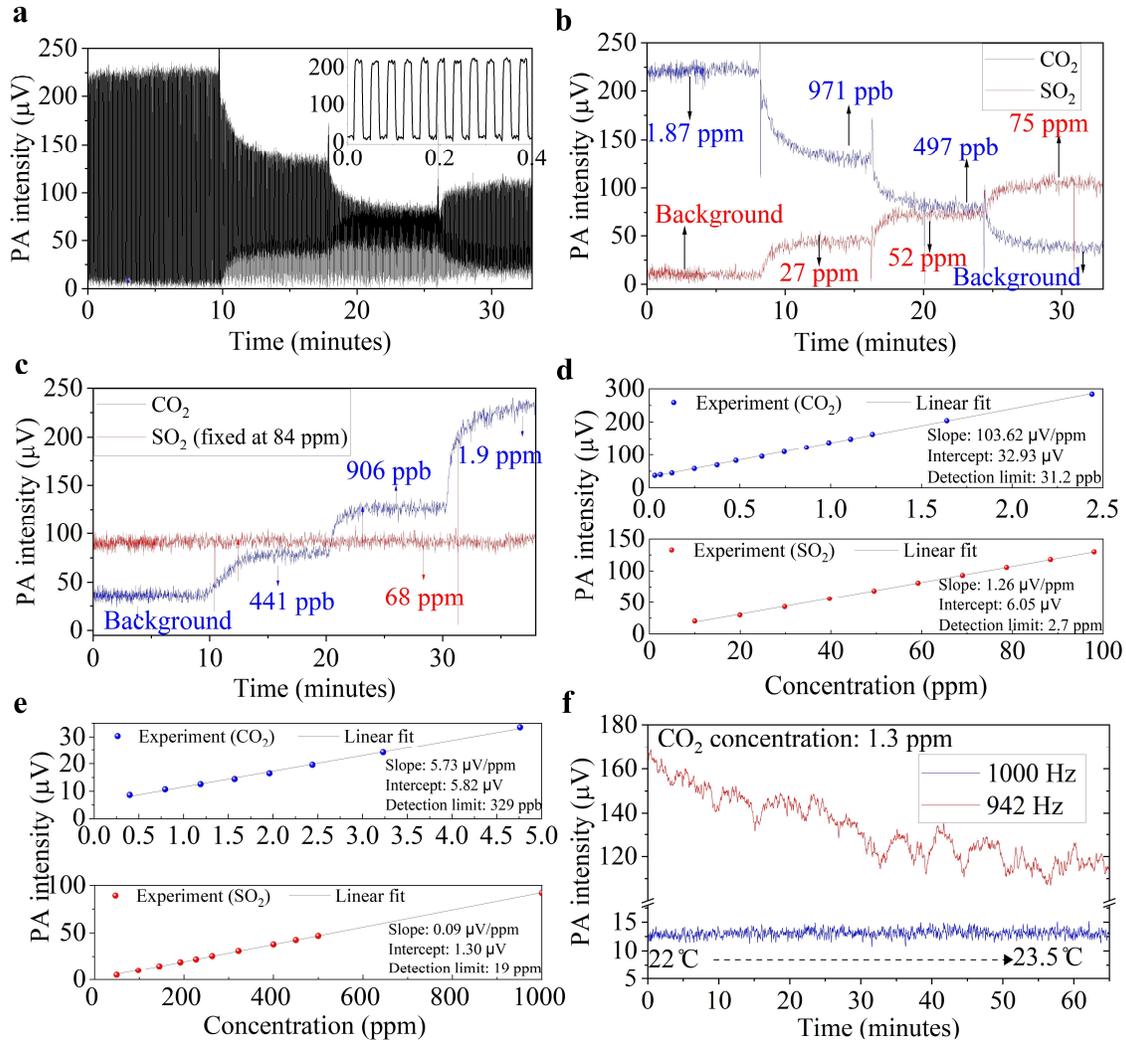

**Fig. 3 | Real-time PA detection of SO$_2$ and CO$_2$ using the proposed gas-filled fiber Raman laser consisting of two synthesized mid-IR spectral lines. a** Raw data stream output from the lock-in amplifier, containing the PA signal excited from both CO$_2$ and SO$_2$ with varied concentrations as a function of time. Inset: a zoom-in result of **a**. **b** PA signals of CO$_2$ and SO$_2$ extracted in real-time from **a**. **c** Real-time PA signal recorded for varied CO$_2$ concentration while constant SO$_2$ concentration at 84 ppm. **d** and **e** are linear fits of the PA intensity versus gas concentrations at different repetition rates of 1.00 and 942 kHz, respectively. **f** The monitoring of PA intensity excited from 1.3 ppm CO$_2$ by increasing the temperature of acoustic chamber from 22 °C to 23.5 °C, using the Raman spectral line at 4.25 μm but with two different repetition rates of 942 Hz and 1 kHz, respectively.

Since the data stream obtained from the lock-in amplifier contains the acoustic information of multiple gases (SO$_2$ and CO$_2$ here), a MATLAB program was developed to separate each other in real-time (see details in Supplementary Note 6). Figure 3b shows the processed result corresponding to Fig. 3a, where the acoustic signals generated by SO$_2$ and CO$_2$ are distinctly separated, allowing for a clear visualization of the evolutionary trend of each gas. A fast-ward video is provided to show this real-time monitoring process (see Supplementary video). Furthermore, as aforementioned, the narrow linewidth and precise tunability of all Raman laser lines bring the advantage of high selectivity and thus low cross-sensitivity. As an example, we carried out another measurement where the $CO_2$ concentration increases in steps while the $SO_2$ concentration is set to a constant value of 84 ppm. Figure 3c shows the measurement result, where one can see that the acoustic intensity of $SO_2$ is indeed independent of the variation of $CO_2$ concentration.

The sensitivity and detection limit of each gas is evaluated by only operating the corresponding Raman laser line and turning off other spectral lines. In this case, the burst form is replaced by periodic Raman pulses in the continuous form. Figure 3d shows the measured PA intensity at different $CO_2$ and $SO_2$ concentrations, respectively, based on the measured PA data in Supplementary Note 7. The laser's repetition rate remains at 942 Hz to include both modulation and pulse regimes of the PA detection. In Fig. 3d, the linear dependence of the PA intensity on the concentration is in accordance with the results being reported [18,34]. The fitted slope (i.e., sensitivity) is 103.6 µV/ppm for $CO_2$ while 1.26 µV/ppm for $SO_2$. The different sensitivity is because the absorption coefficient of $CO_2$ at 4.25 µm is around two orders of magnitude higher than $SO_2$ at 3.99 µm (see Fig. 2c). As a comparison, Fig. 3e presents the measured results with the same setting as Fig. 3d but using 1 kHz repetition rate, where its 4[th] order harmonic is far away from any resonant acoustic peaks in Fig. 2f. In this case, the PA detection operates only in the pulse regime, and the sensitivity decreases to 5.7 µV/ppm for $CO_2$ and 0.09 µV/ppm for $SO_2$ (see Fig. 2c). Despite this, the PA detection operating at 1 kHz repetition rate brings the important advantage of high tolerance on the fluctuation of environmental temperature and is thus more robust towards real-world applications, because it excludes the contribution of the modulation regime which is sensitive to temperature [18]. To validate this statement, we monitored the PA intensity of $CO_2$ at 1.3 ppm concentration over 65 minutes when the acoustic chamber was heated up from 22 to 23.5 °C (see details in Supplementary Note 8). As shown in Fig. 3f, the PA intensity does maintain a good stability at 1 kHz repetition rate, but significantly decreases over time at 942 Hz.

The detection limit is estimated using the "propagation of errors model" in Ref. 42, which incorporates the errors in the analyte measurements including the blank sample. The calculation is based on the measured data points in Fig. 3d and 3e as well as the standard deviation and mean value of the blank sample provided in Supplementary Note 9. The detection limit is calculated to be 31.2 ppb for $CO_2$ and 2.7 ppm for $SO_2$ at 942 Hz repetition rate, while 329 ppb for $CO_2$ and 19 ppm for $SO_2$ at a 1 kHz repetition rate. Given the measured 3.24 $m^{-1}$ attenuation coefficient of the 4.25 μm Raman laser in the ambient air with ~400 ppm concentration (see Supplementary Fig. S5c) and 105 mHz noise equivalent bandwidth of the lock-in amplifier, the normalized noise equivalent absorption (NNEA) coefficient is estimated to be $1.1\times10^{-4}$ $W.cm^{-1}.Hz^{-1/2}$ at 942 Hz and $1.2 \times10^{-3}$ $W.cm^{-1}.Hz^{-1/2}$ at 1 kHz repetition rate. The NNEA for $SO_2$ detection cannot be estimated because its absorption coefficient is too weak to be measured at ppm concentration in this experiment.

## Discussion

In our investigation, setting a longer time constant for the lock-in amplifier can further mitigate the fluctuation of the measured acoustic data and thus can reach a lower detection limit. This statement is supported by the Allan deviation shown in Fig. 4 which is calculated from the background acoustic signal shown in Supplementary Note 9. The Allan deviation keep decreasing toward a longer integration time, indicating that a lower detection limit can be achieved by setting a longer time constant for the lock-in amplifier [43,44].

However, it should be noted that, using a longer time constant prolongs the response time of detecting multiple gases, thereby compromising the proposed concept of "real-time detection". An alternative solution is to enhance the sensitivity of the PA detection by further narrowing the Raman laser's linewidth and boosting its pulse energy because it can further enhance the intensity of the excited acoustic waves by concentrating more laser line energy into one of the absorption lines of the selected gas. Furthermore, a narrow linewidth can suppress the cross-sensitivity and thus achieve high accuracy/selectivity in absorption-based gas spectroscopy because the absorption spectra of the target gases typically overlap with each other and other possible background gases. In this work, the linewidth of the mid-IR Raman laser from the 2$^{nd}$ stage $H_2$-filled ARHCF is measured to be ~1.3 nm. Since $CH_4$ filled in the 1$^{st}$ ARHCF is known with a relatively large vibrational Raman gain width (~600 pm at 2 bar) [32,45], the primary factor for further compression of the laser's linewidth is to replace $CH_4$ with another gas with a narrower Raman gain width. An example is to use the rovibrational SRS effect at the Q-branch of the $v_1$ band of the $v_1/2v_2$ Fermi dyad of $CO_2$ which has a much narrower Raman gain width (few pm at 2 bar) [31,46,47]. The feasibility of using $CO_2$ is initially supported by the measured spectra presented in Fig. S1b, but the wavelength selection of Raman lines needs to take the disturbance of infrared absorption bands of $CO_2$ into consideration. Reducing the gas pressure can narrow the linewidth of the Raman laser as well. However, gas pressure reduction is associated with a decrease in the Raman gain coefficient, and thus requires further reduction on the optical loss of the cascaded ARHCFs. Narrowing the pump linewidth is an additional option, but it involves the suppression of the SBS effect in the Yb-doped fiber amplification stages.

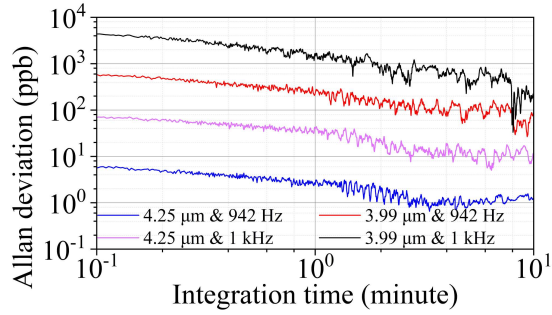

**Fig. 4 | Allan deviations of the background PA signals as a function of averaging time at different repetition rates and wavelengths of the proposed gas-filled fiber Raman laser.**

The quasi-linear decrease of the Allan deviation in logarithmic scale in Fig. 4 is also an indication of the good long-term stability of the proposed laser-PA system, ensuring good repeatability of the gas detection, which has been experimentally validated in Supplementary Note 10.

The proposed laser is also adaptive to other absorption-based gas detection modalities. For example, it can be combined with a multi-pass cell for conducting direct absorption gas spectroscopy in the time domain, where the gas concentration information can be extracted from the amount of Raman pulse energy/peak power being absorbed by the gases. In this case, the switching time from one to another gas species will be as short as a few millimeters, owing to the rapid switching between different Raman laser lines, as illustrated in Fig. 2b.

In summary, we have proposed a novel concept for the synthesis of multiple narrow and intense spectral lines over a broad near- and mid-IR range. The suggested concept provides the possibility of reconfiguring the number and wavelengths of all spectral lines according to the targeting gases. The proposed approach is expected to push the state-of-the-art in multi-gas detection a step forward in terms of selectivity, sensitivity, and response time. A gas-filled fiber Raman laser synthesizing two independently and precisely tunable narrow spectral lines at 3.99 and 4.25 μm is developed and used as a proof-of-concept for the PA detection of $SO_2$ and $CO_2$ in the ppm and ppb concentration range. The proposed concept opens the way for the development of promising light sources within a wide range of applications including IR trace gas spectroscopy and multi-spectral microscopy, while having the potential to be developed into a compact and robust all-fiber structure [48–50].

## Methods
### Optical amplification module of the pump laser

The 1044/1060 nm laser diode seeds being combined into a single PM fiber for optical amplification have the same linear polarization direction aligned to one of the birefringence axes of the PM Yb-doped fiber amplifiers. The amplification module adopts two stages pre-amplifiers and a power amplifier. The core diameters of Yb-doped active PM fibers used in the $1^{st}$-$3^{rd}$ stage amplifiers are 6 μm (PM-YSF-HI-HP, Coherent), 10 μm (PLMA-YDF-10/125, Coherent), and 25 μm (PLMA-YDF-25/250, Coherent), respectively. Optical isolators are used after each amplification stage, to block backward detrimental light. The output fiber of the power amplifier is angle-cleaved to mitigate back-reflection. 915 nm laser diodes are used as pumps. They are directly modulated to emit pulses with width > 50 μs, to support the pulse pump regime. All pulsed laser diode pumps are synchronized to provide the required energy for the amplification of the signal laser pulses.

The pulse width of the seed laser is optimized to 3.7 ns. The detrimental SBS effect becomes significant during the amplification when the signal pulse width > ~4 ns, although a long pulse width is preferable in term of suppressing the transient Raman regime and enhancing the efficiency of the $2^{nd}$ stage $H_2$-filled ARHCF Raman laser generation.

**Development of the gas-filled ARHCF**

The ARHCF is fabricated with the stack-and-draw method. Custom-made high-pressure gas cells were developed to seal the ARHCFs for gas filling and light coupling. In the 1-30 bar pressure range, the test shows that the gas cell has an excellent gas sealing ability (leakage speed is measured to be ~0.01 bar/hour at 30 bar, and ~$10^{-4}$ bar/hour at 2 bar), which ensures the Raman laser's long-term stability. Gases that are used for the ARHCF setup for Raman laser generation have a high purity of >99.9999%.

**Photoacoustic system**

The cross-section of the acoustic chamber is shown at the bottom of Fig. 2a. The main body of the acoustic chamber is a hollow-core cylinder with a 5.15 cm internal radius and 10.3 cm length [38,39]. Acoustic filter elements are mounted at each end of the chamber, to suppress ambient acoustic noise at few resonant acoustic frequencies. All components of the acoustic chamber are made of stainless steel, and their inner surfaces are well-polished to ensure the uniformity of the acoustic reflections. The two ends of the acoustic chamber are equipped with two sapphire optical windows with an anti-reflection coating at the mid-IR region, to provide low-loss access for the developed Raman spectral lines. A 1/2" condenser low-noise microphone (4955, Brüel & Kjær) is placed halfway between the ends of the acoustic cylinder for recording the excited acoustic waves. The front surface of the microphone is aligned with the inner surface of the cylinder. The microphone has an integrated low-noise amplifier, which is followed by a conditioning amplifier with the functions of both supplying power to the microphone and amplifying the acoustic signal being recorded (1708, Brüel & Kjær). The final output acoustic data from the microphone is recorded by either a high-speed digitizer (M4i4421-x8, SPECTRUM) or a lock-in amplifier (MFLI 500 kHz, Zurich Instruments).

The preparation of the gas sample is based on commercial pressurized gas products including 100 ppm $CO_2$, pure $CO_2$, 100 ppm $SO_2$, 5000 ppm $SO_2$, and pure Ar (Air Liquide A/S). Ar is used as the gas diluent. MFCs calibrated with pure Ar are used to regulate the flow rate of these commercial gases. Gases output from all MFCs are combined to a single gas hose, which is connected to a gas chamber to uniformly mix the concentrations. The gas concentration is regulated by controlling the flow rate with the MFCs. The gas chamber's output is directed into a T-shaped hose splitter, to divide the gas flow into two ports, A and B. Port B is linked to the laboratory's ventilation point through a needle valve, which serves to control the flow rate in port A. The flow rate in port A is monitored by a mass flow meter (MFM), and then the MFM directs the prepared gas sample into the acoustic chamber via a gas inlet port, where the gas sample flows through the main body of the acoustic chamber and exits through a gas outlet. In this experiment, the gas flow rate is <= 300 sccm, to mitigate the noise rising from the turbulence of the gas flow.

**Acknowledgements.** This work is supported by the Danmarks Frie Forskningsfond Hi-SPEC project (Grant No. 8022-00091B), VILLUM Fonden (Grant No. 36063, Grant No. 40964), LUNDBECK Fonden (Grant No. R346-2020-1924), and US ARO (Grant No. W911NF-19-1-0426). We thank Martin Nielsen (affiliated with DTU Space) for fabricating the gas cells and the PA chamber.

**Competing interests.** The authors declare no conflicts of interest.

**Author contributions.** Y. W. and C. M. conceived the concept. Y. W. designed and developed the whole experiment system under the assistance of C. M., M. K. D. and A. I. A.. J. E. A., J. W and R. A. C. designed and fabricated ARHCFs. C. M. measured the scanning electron microscope images of ARHCFs. Y. W. and L. H. conducted the gas detection experiments. Y. W. performed the data analysis and conducted the loss simulation of ARHCFs using COMSOL software. Y. W prepared the figures and manuscript under the discussion with C. M., M. K. D. and A. I. A.. C. Z. improved the figure quality and prepared the manuscript format. All the authors contributed to the improvement of manuscript quality.

**Supplementary Information for "Synthesizing gas-filled fiber Raman lines enables access to the molecular fingerprint region"**

Yazhou Wang, Lujun Hong, Cuiling Zhang, Joseph Wahlen, J. E. Antonio-Lopez, Manoj K. Dasa, Abubakar I. Adamu, Rodrigo Amezcua-Correa and Christos Markos

# SUPPLEMENTARY NOTE 1. Examples supporting the proposed concept

**Supplementary Table 1. Examples showing the wavelength range of Raman Stokes lines using different gases.** VSRS: vibrational stimulated Raman scattering. RSRS: rotational stimulated Raman scattering.

| | 1st stage gas | | 2nd stage gas | |
|---|---|---|---|---|
| | Gas species and SRS type | Wavelength | Gas species and SRS type | Wavelength |
| Pump lines' range (1015-1115 nm) | 1st order VSRS of $CH_4$ (2917 cm$^{-1}$) | 1442-1652 nm | 1st order VSRS of $H_2$ (4155 cm$^{-1}$) | 3597-5268 nm |
| | 1st order VSRS of $CO_2$ (1385 cm$^{-1}$) | 1181-1319 nm | 1st order VSRS of $H_2$ (4155 cm$^{-1}$) | 2319-2918 nm |
| | 2nd order VSRS of $CO_2$ (1385 cm$^{-1}$) | 1412-1614 nm | 1st order VSRS of $H_2$ (4155 cm$^{-1}$) | 3416-4900 nm |
| | 1st order VSRS of $N_2$ (2331cm$^{-1}$) | 1330-1507 nm | 1st order VSRS of $H_2$ (4155 cm$^{-1}$) | 2973-4031 nm |

The wavelength conversion examples in Supplementary Table 1 are used for showing the proposed concept reflected by Equation (1) in the main text. The range of the pump lines is set as 1015 to 1115 nm, which is the typical Yb-doped fiber gain range [1]. From this table, it can be seen that the wavelength range of Raman Stokes is much larger than the pump wavelength range. For example, by using the VSRS effect of $CH_4$ and $H_2$, the Raman Stokes wavelength range is spanning from 3597 to 5268 nm, which is ~15 times of the pump range. Note that, under the framework of gas-filled ARHCFs, the wavelength range is also determined by the transmission window of ARHCF. The maximum wavelength of silica ARHCF-based laser is supposed to be 4.6 μm because of the dramatic increase of silica glass absorption loss towards longer wavelength [2].

It is worth noting that, to develop a Yb-doped fiber amplifier covering the full gain range from 1015 to 1115 nm, parallel amplification stages with complementary working bands (e.g., 1015 – 1055 nm, and 1055 - 1115 nm) need to be adopted. This is mainly because the gain wavelength range of Yb-doped fiber shifts as a function of gain fiber length due to the re-absorption effect. Output lasers from these parallel amplification stages can be combined into a single fiber, to pump subsequent gas-filled cascaded ARHCFs whose details are shown in Supplementary Note 2.

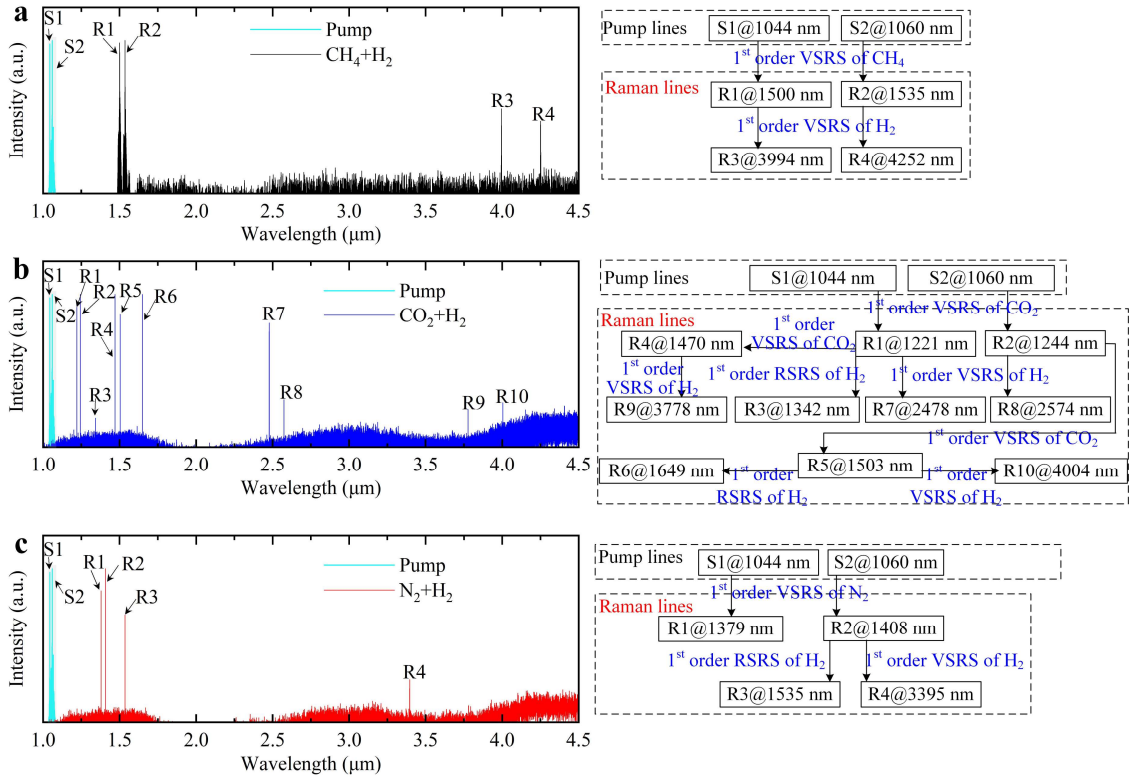

**Fig. S1 | Raman lines measured from the output of the 2$^{nd}$ stage H$_2$-filled ARHCF through filling different gases into the 1$^{st}$ stage ARHCF.** The block diagrams in right side shows the relations of wavelength conversion for its corresponding spectral lines in the left side.

The left side of Fig. S1 presents the measured Raman lines using the wavelength conversion schemes in Supplementary Table 1, where the pump consists of two narrow spectral lines at 1044 and 1060 nm. The wavelength conversion process in each case is elaborated in the block diagrams shown on the right side of Fig. S1. Supplementary Table 2 shows the summary of measured wavelengths of Raman line pairs by following the wavelength conversion schemes in Supplementary Table 1. It can be seen that the line spacing of the Raman line pairs output from the 2$^{nd}$ stage H$_2$-filled ARHCF is much larger than the 16 nm line spacing of the pump lines. Note that the line spacing is missing in the case of using N$_2$ for the 1$^{st}$ stage ARHCF, because the energy of the R1 line @ 1379 nm is not sufficient to generate the mid-IR vibrational Raman line through the VSRS effect of the 2$^{nd}$ stage H$_2$-filled ARHCF.

**Supplementary Table 2. Measured wavelengths of Raman lines by following the wavelength conversion schemes in Supplementary Table 1.** Note that rotational Raman lines are not considered in this table. Line spacing in this table refers to the wavelength spacing of the pair of mid-IR Raman lines output from the 2$^{nd}$ stage ARHCF.

| | 1$^{st}$ stage gas-filled ARHCF | | 2$^{nd}$ stage H$_2$-filled ARHCF | | Line spacing |
|---|---|---|---|---|---|
| | Gas species and SRS type | Wavelengths | Gas species and SRS type | Wavelength | |
| Pump lines' range (1044&1060 nm) | 1$^{st}$ order VSRS of CH$_4$ (2917 cm$^{-1}$) | R1@1500 & R2@1535 nm | 1$^{st}$ order VSRS of H$_2$ (4155 cm$^{-1}$) | R3@3994 & R4@4252 nm | 258 nm |
| | 1$^{st}$ order VSRS of CO$_2$ (1385 cm$^{-1}$) | R1@1221 & R2@1244 nm | 1$^{st}$ order VSRS of H$_2$ (4155 cm$^{-1}$) | R7@2478 & R8@2574 nm | 96 nm |
| | 2$^{nd}$ order VSRS of CO$_2$ (1385 cm$^{-1}$) | R4@1470 & R5@1503 nm | 1$^{st}$ order VSRS of H$_2$ (4155 cm$^{-1}$) | R9@3778 & R10@4004 nm | 226 nm |
| | 1$^{st}$ order VSRS of N$_2$ (2331cm$^{-1}$) | R1@1379 & R2@1408 nm | 1$^{st}$ order VSRS of H$_2$ (4155 cm$^{-1}$) | unknown & R4@3395 nm | unknown |

The spectra in Fig. S1 were measured using two different spectrum analyzers. The spectrum in Fig. S1a was measured using a scanning-grating-based spectrum analyzer (Spectro 320, Instrument Systems). In this case, the pump laser operates in the regime of dual spectral lines, i.e., the single spectrum consisting of all Raman lines. In Fig. S1b and S1c, the spectra were measured using a Fourier transform optical spectrum analyzer (OSA207C, Thorlabs). In this case, each spectrum was obtained by running either of the dual pump spectral lines, to obtain a sufficient signal-to-noise ratio. Therefore, in Fig. S1b and S1c, there are multiple measured spectra overlapped to show all Raman lines. Note that the raw spectra measured from these two instruments have different sensitivities and background intensities. To offset the mismatch, these measured spectra are properly shifted, stretched, and normalized along the intensity direction. This adjustment facilitates the visualization in Fig. 1b in the main text.

# SUPPLEMENTARY NOTE 2. Characterization of ARHCFs

As shown in Fig. S2a, the 1st stage ARHCF has a 32.8 μm core diameter surrounded by seven silica capillaries, where each capillary has a diameter of 16.1 μm and a wall thickness of 323 nm. The 2nd stage ARHCF has an 82.0 μm core diameter, and the cladding is composed of seven nested silica capillaries for suppressing the bending loss [3]. The diameter and wall thickness are 40.3 μm and 987 nm for the external capillary, and 13.6 μm and 1.37 μm for the internal capillary, respectively.

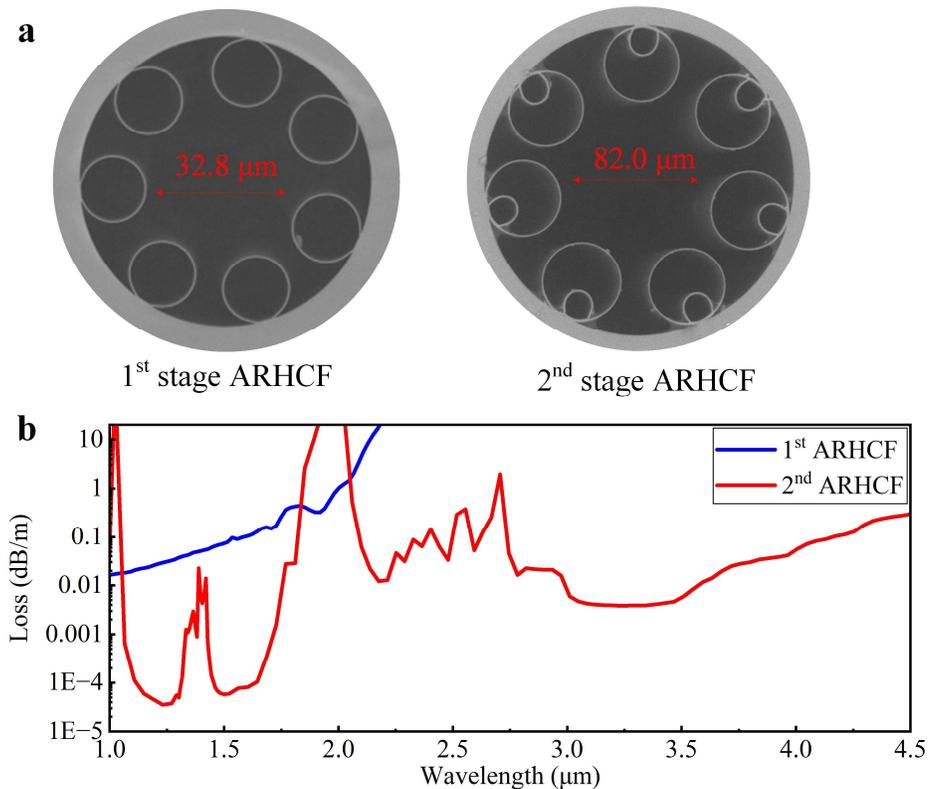

**Fig. S2 | 1st and 2nd stage ARHCFs. a** Scanning electron microscope images. **b** Simulated loss spectra.

The simulation was implemented based on the finite-element method using the COMSOL software with parameter settings as Ref. 4, but without taking into consideration the surface scattering loss because of the negligible surface roughness of the silica ARHCF compared to the wavelength in the IR region [3].

The simulation result in Fig. S2b shows that the 1st stage ARHCF has a low loss < 0.1 dB/m in the wavelength range of 1.0-1.6 μm region while a high loss towards longer wavelength, which blocks the generation of 2$^{nd}$ order vibrational Raman Stokes of $CH_4$ at ~2.7 µm and thereby boosts the energy of 1$^{st}$ order vibrational Raman Stokes. The 2$^{nd}$ stage ARHCF shows mainly two low-loss windows separated by a strong resonance peak at ~2 µm, ensuring the generation of multiple Raman laser lines over a broad IR range.

# SUPPLEMENTARY NOTE 3. Characterization of the pump spectral lines

The synthesized pump spectral lines from the Yb-doped fiber amplifier are characterized at their maximum energies. An optical spectrum analyzer (ANDO AQ6317B, AssetRelay) with a resolution of 0.01 nm is used to measure the optical spectrum. Figure S3a shows the optical spectrum measured when S1 and S2 alternately output in the time domain, where it presents dual spectral lines at 1044 and 1060 nm. The Gaussian-like beam profile (measured by a beam profiler (BP109-IR2, Thorlabs)) in the inset of Fig. S3a indicates that the laser operates in the fundamental mode of the fiber. The spectrum of each line is also individually measured at different temperatures of the corresponding laser seed, as shown in Fig. S3b. Their linewidths are measured to be ~0.13 nm. Each line can be thermally tuned over the ~1 nm range, and the wavelength is linearly dependent on the temperature (see Fig. S3c). The pulse widths of both lines are measured to be ~3.7 ns by an InGaAs photodetector (5 GHz bandwidth, DET08C, Thorlabs) connecting to an oscilloscope (6 GHz bandwidth, MSO64B, Tektronix) (see Fig. S3d). The relative intensity noise (RIN) is measured at the linear response region of the photodetector. Based on 10,000 pulses, the RIN is measured to be 0.7% at 1044 nm and 0.8% at 1060nm (see Fig. S3e). Figure S3f and S3f g shows the long-term stability of the laser in terms of the pulse peak intensity (measured by photodetector) and pulse energy (measured by energy meter), respectively. The long-term stability is measured after a three-hour warming-up of the laser system. The pulse energies in Fig. S3g are measured to be ~75 µJ @ 1044 nm and ~98 µJ @ 1060 nm, respectively. Insets of Fig. S3f and S3g show the zoom-in alternate output between these two spectral lines.

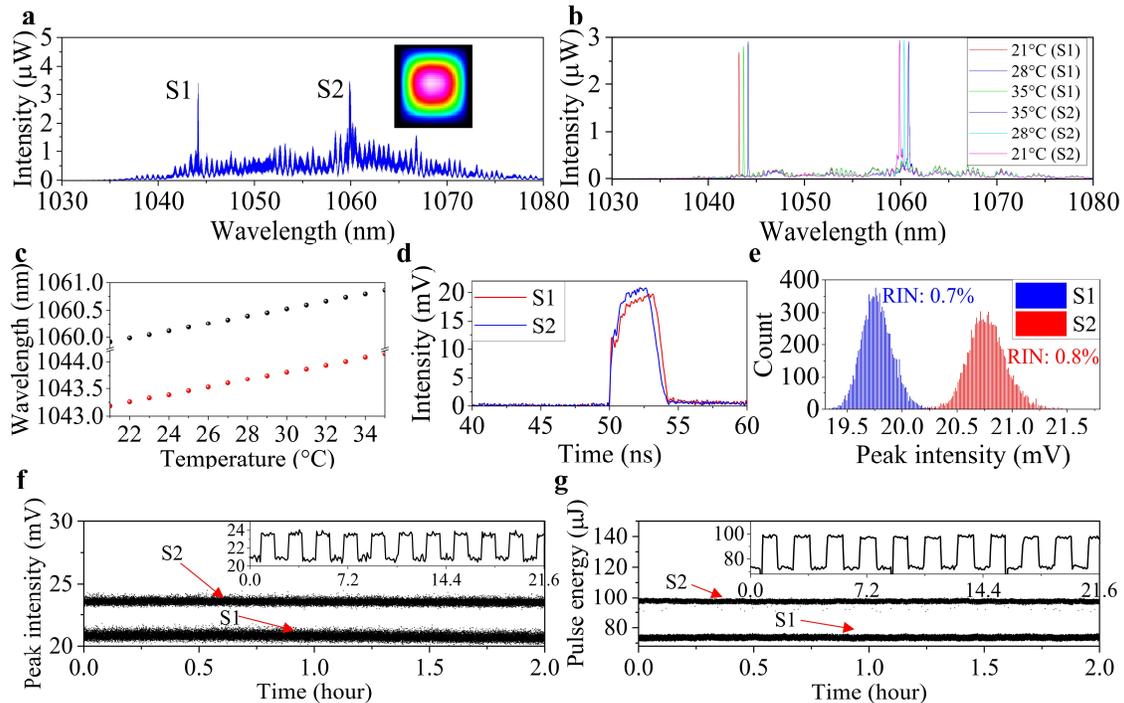

**Fig. S3 | Characterization of the pump laser synthesized of two spectral lines at 1044 and 1060 nm. a** Measured optical spectrum. Inset is the beam profile of S2. S1 and S2 refers to seed1 at 1044 nm and seed2 at 1060 nm, respectively. **b** Optical spectra at single wavelength operation with different temperature for laser diode seeds. **c** Thermal tuning of wavelengths of laser diode seeds. **d** Pulse profiles. **d** Histograms of measured pulse peak intensity. **f** and **g** are the monitoring results (scattering graph) of pulse peak intensity and pulse energy over 2 hours measured by a photodiode and an energy meter, respectively. Insets of **f** and **g** are zoom-in (line graph) of the corresponding results.

# SUPPLEMENTARY NOTE 4. Characterization of the 1st stage CH$_4$-filled ARHCF Raman Stokes lines

The 1st order vibrational Raman Stokes spectral lines are generated from the 1st stage CH$_4$-filled ARHCF pumped by the Yb-doped fiber laser with synthesized spectral lines at 1044 and 1060 nm. Figure S4a shows measured spectra, where the spectrum of each line is individually measured by turning off the other line. Their linewidths are measured to be ~0.5 nm by the optical spectrum analyzer (ANDO AQ6317B, AssetRelay) with a resolution of 0.01 nm. The Gaussian-like beam profile in the inset of Fig. S4a indicates that the Raman laser operates in the fundamental mode of the fiber. The 2nd order Raman Stokes at 2.7 μm was not observed because of the high fiber loss. Figure S4b shows the wavelength tuning range of each Raman spectral line through thermally tuning its corresponding pump line (see Fig. S4b and S4c). The tuning range is from 1500.0 to 1501.9 nm for the 1st Raman line, and from 1534.7 to 1536.7 nm for the 2nd Raman line. Figure S4c shows the measured pulse profiles by a photodetector (5GHz bandwidth, DET08C, Thorlabs), where their pulse widths are measured to be ~3.3 ns for both wavelengths. The RIN is measured to be 5.5% and 4.9 %, which are higher than the RIN of the pump lines due to the inherent quantum noise during the SRS process [5].

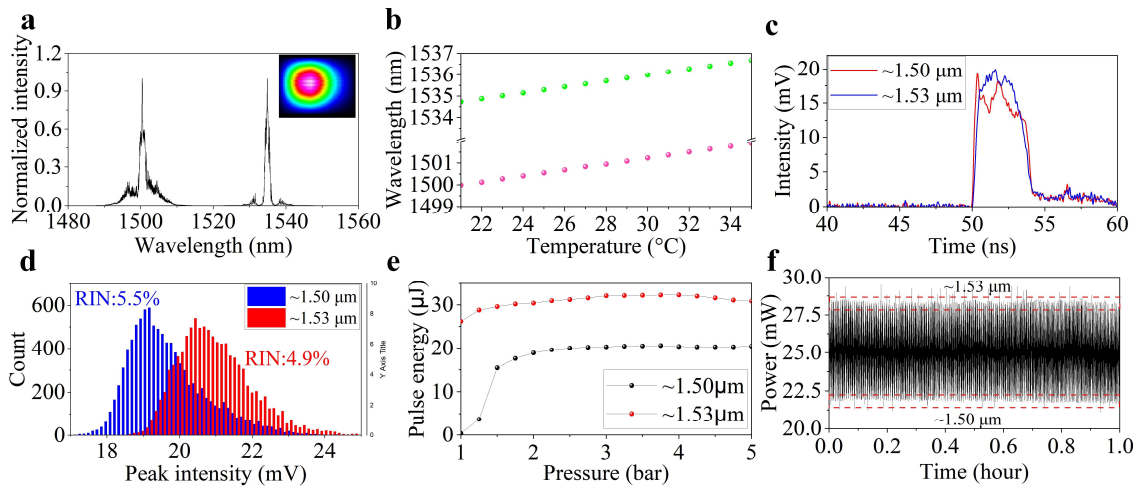

**Fig. S4 | Characterization of the synthesized Raman spectral lines from the CH$_4$-filled 1st stage ARHCF. a** Measured optical spectra. Inset is the beam profile of 1.53 μm Raman laser. **b** Wavelength tunability by thermally tuning the pump laser's wavelength. **c** Pulse profiles. **d** Histograms of measured pulse peak intensity. **e** Measured pulse energies of 1.50 and 1.53 μm Raman laser lines as a function of CH$_4$ pressure. **f** Average power monitoring over 1 hour.

The Raman pulse energy is optimized in terms of $CH_4$'s pressure from 1 bar to 5 bar, as shown in Fig. S4e. With the increase of gas pressure, the Raman pulse energy increases and gradually approaches a saturation level due to the known saturation effect of the Raman gain [6]. At 2.5 bar pressure, the pulse energies are measured to be 30 µJ at 1.50 µm and 22 µJ at 1.53 µm, respectively. Figure S4f shows the measured average power when the two Raman spectral lines alternately output, where each laser burst is composed of 1000 pulses with 1 kHz internal repetition rate.

# SUPPLEMENTARY NOTE 5. Characterization of the 2nd stage $H_2$-filled ARHCF Raman Stokes lines

The 1st stage $CH_4$-filled ARHCF Raman laser synthesized of 1.50 and 1.53 μm spectral lines is used to pump the 2nd stage $H_2$-filled ARHCF, to generate synthesized vibrational Raman Stokes lines at 3.99 and 4.25 μm. Figure S5a shows the measured pulse energy as a function of $H_2$ pressure. The pulse energy increases towards high pressure due to the increase of Raman gain coefficient, and the maximum pulse energy is measured to be 1.7 μJ at 3.99 μm and 2.7 μJ at 4.25 μm, respectively, at the $H_2$ pressure of 30 bar. The evolution trend in Fig. S5a implies that higher pulse energy can be obtained by further increasing the gas pressure, however, the maximum pressure for our gas-filled ARHCF system is designed to be 30 bar (see Methods). The pulse energy at 4.25 μm is lower than 3.99 μm because of the absorption of the ~400 ppm concentration $CO_2$ in ambient air. This statement is supported by a power attenuation measurement in ambient air, as shown in Fig. S5c, where the Raman power at 4.25 μm is measured as a function of propagation distance at two different seed temperatures of 22 and

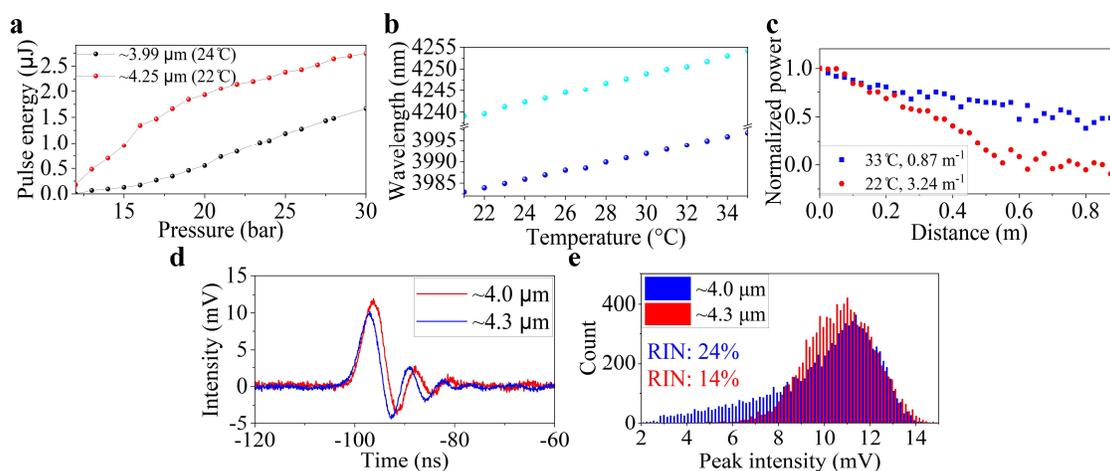

**Fig. S5 | Characterization of the synthesized two mid-IR Raman spectral lines from the $H_2$-filled 2nd stage ARHCF. a** Measured pulse energies of 3.99 and 4.25 μm Raman Stokes lines as a function of $CH_4$ pressure. **b** Wavelength tunability by thermally tuning the pump laser's wavelength. **c** Attenuation of 4.25 μm Raman laser as a function of propagation distance in the ambient air under different seed laser diode's temperatures. **d** Pulse profiles. **e** Histograms of measured pulse peak intensities.

33 °C. The Raman laser beam is collimated during the measurement, to ensure that the power is completely detected by the power meter. The attenuation coefficients are measured to be 3.24 m$^{-1}$ at 22 °C and 0.87 m$^{-1}$ at 33 °C, respectively. Compared with 33 °C, the higher attenuation coefficient at 22 °C is consistent with the spectra presented in Fig. 2c in the main text, where the Raman line at lower temperature overlaps with the stronger absorption line of $CO_2$.

The pulse profiles in Fig. S5d are measured using a HgCdTe amplified photodetector (PDAVJ8, 100 MHz bandwidth, Thorlabs). In this case, due to the limited bandwidth of the photodetector, the pulse shows a negative part, and the pulse width cannot be accurately measured. The RIN is measured to be 24% at 3.99 μm and 14% at 4.25 μm, respectively. The lower RIN at 4.25 μm is also a sign of the higher Raman pulse energy when compared to 3.99 μm [5]. Despite the relatively high RIN, the PA signal in Fig. 3a in the main text shows a small fluctuation because of the average function of the lock-in algorithm. Lower RIN is anticipated by further boosting the Raman pulse energy through optimizations [5].

# SUPPLEMENTARY NOTE 6. Processing of the acoustic data excited from multiple gases using the proposed gas-filled fiber laser source

With the proposed concept, the acoustic data stream acquired from the signal output port of the lock-in amplifier contains acoustic information excited from multiple gases. The data is transferred to a computer in real-time by using the "LabOne" software (Zurich Instruments), which provides a MATLAB application program interface. The real-time separation of the acoustic information of each gas is achieved with the assistance of a high-speed digitizer (M4i4421-x8, SPECTRUM) synchronizing the trigger message of laser diode seeds at 1045 and 1060 nm. Two channels of the digitizer are used to receive the trigger information from the laser diode seeds at 1044 and 1060 nm, respectively, which provide the reference for the separation of acoustic signal excited from different gas species. A MATLAB code is developed for real-time processing the acoustic data stream by synchronizing the lock-in amplifier and the digitizer.

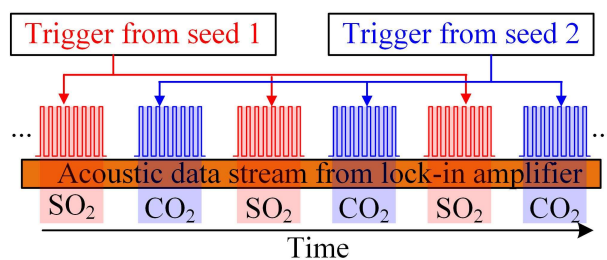

**Fig. S6 | Illustration of the real-time data processing method for the multi-gas PA data stream output from lock-in amplifier**.

Figure S6 shows the principle of the code. The acoustic data stream output from the lock-in amplifier is continuously read by the MATLAB code. The stream occurring in the period of the trigger burst of seed1/seed2 is extracted as the acoustic data excited from $SO_2$/$CO_2$, respectively.

# SUPPLEMENTARY NOTE 7. Raw photoacoustic data for the sensitivity measurement of gas detection

Figure S7 shows the evolution of the recorded PA signal at different concentrations. In this measurement, the time constant and order of the low-pass filter of the lock-in amplifier are set to 500 ms and 5, respectively. According to the target gas, the Raman laser operates at either 4.25 μm or 3.99 μm wavelength. The gas flow is regulated to be less than 300 sccm by using MCFs and needle valve. To prepare Fig. 3d and 3e in the main text, the PA intensity at each concentration is calculated as the average of the corresponding stable region of the step-like evolution data in Fig. S7.

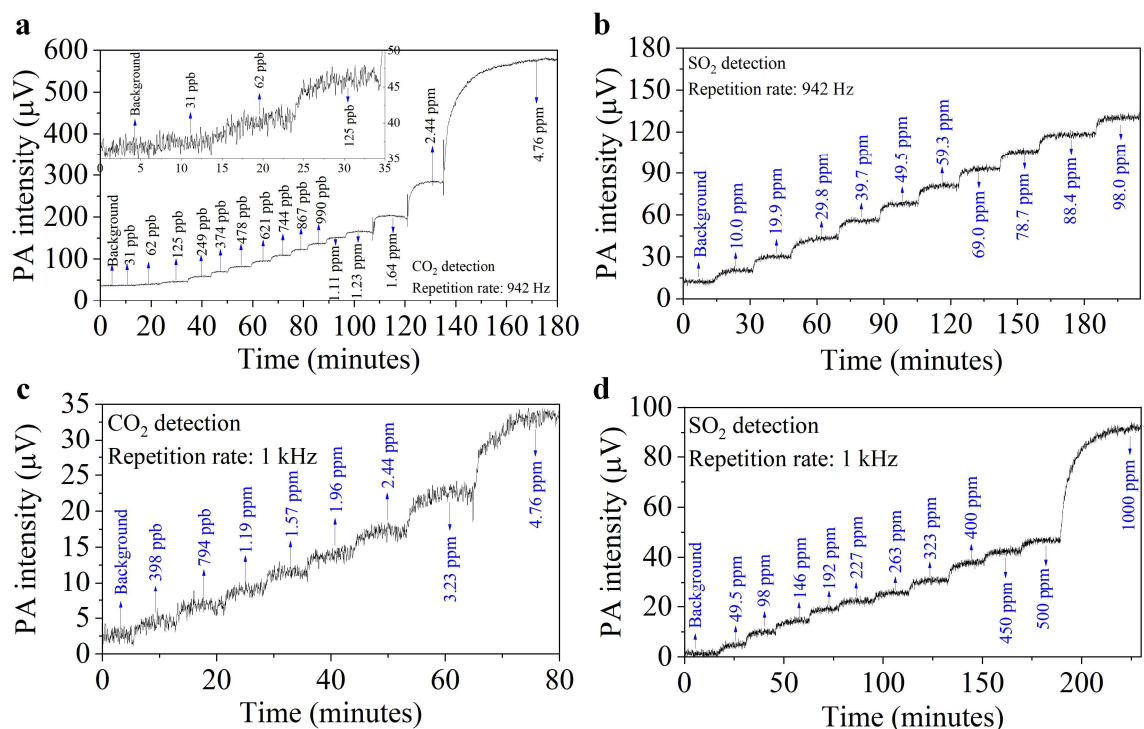

**Fig. S7 | PA intensity evolution at different gas concentrations and different laser repetition rates.** **a** and **b** are PA intensities recorded at 942 Hz laser repetition rate, while **c** and **d** are at 1 kHz repetition rate. The acoustic data is excited from $CO_2$ in **a** and **c**, while is excited from $SO_2$ in **b** and **d**.

# SUPPLEMENTARY NOTE 8. Heating up the acoustic chamber

The temperature of the photoacoustic chamber is increased by placing a heat source along its side, as shown in Fig. S8. The main part of the heat source is a heat plate whose temperature can be adjusted. In this measurement, the temperature of the heat plate is set to 70 °C, to heat up the acoustic chamber through thermal diffusion. A digital thermometer is used to monitor the temperature of the acoustic chamber through a thermocouple. The temperature of the acoustic chamber increases from 22.0 to 23.5 °C over ~1 hour heating up.

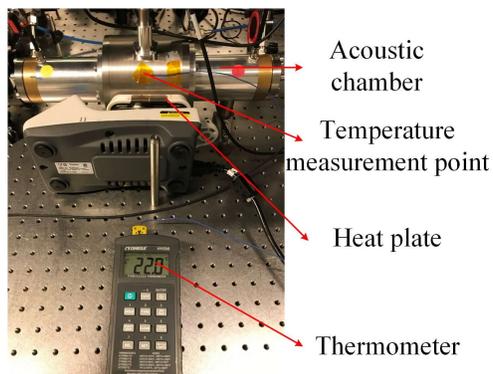

**Fig. S8 | The setup for heating up the acoustic chamber**.

# SUPPLEMENTARY NOTE 9. Raw photoacoustic data for the estimation of gas detection limit

The blank PA data in Fig. S9 was recorded at different Raman laser parameters while the acoustic chamber was constantly flushed with pure Ar at 300 sccm flow rate. The seed temperature is set to 22 °C for the 3.99 µm Raman laser and 24 °C for the 4.25 µm Raman laser. These data are used for:

i) estimating the gas detection limit based on its standard deviation and mean value combined with the calibration line in Fig. 3d and 3e in the main text.

ii) calculating Allan deviations in Fig. 4 in the main text. In this case, the data in Fig. S9 is converted to gas concentration using the calibration lines in Fig. 3d and 3e in the main text.

In this measurement, the time constant and order of the low-pass filter of the lock-in amplifier are set to 500 ms and 5, respectively, which is consistent with Fig. S7.

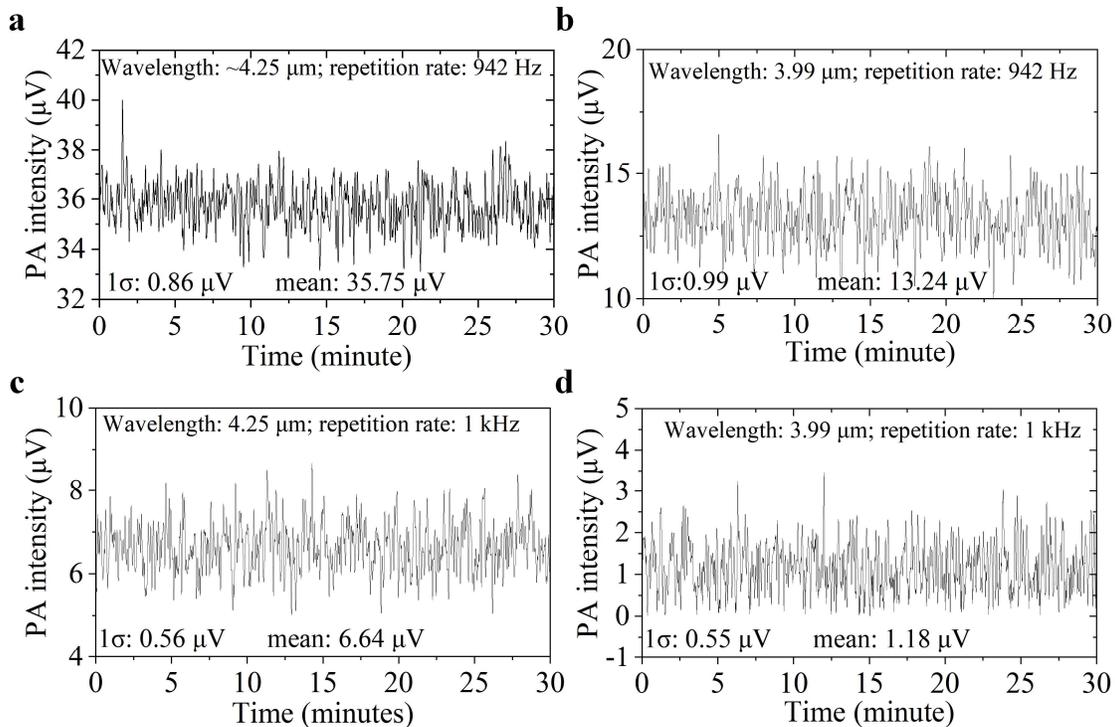

**Fig. S9 | Measured blank PA signal when the Raman laser operates at different wavelengths (3.99 and 4.25 µm) and different repetition rates (942 Hz and 1 kHz).** The data was recorded when the acoustic chamber was flushed with pure Ar.

# SUPPLEMENTARY NOTE 10: Repeatability measurement of the photoacoustic detection of CO$_2$

In this measurement, the repetition rate of the 4.25 μm Raman laser is 1 kHz. The time constant and order of the low-pass voltage filter of the lock-in amplifier are set to 500 ms and 5, respectively. The CO$_2$ concentration is in order set to 0, 910 ppb, 1.9 ppm, 3.8 ppm, 1.9 ppm, 910 ppb, and 0, to record the repeatability of PA detection shown in Fig. S10.

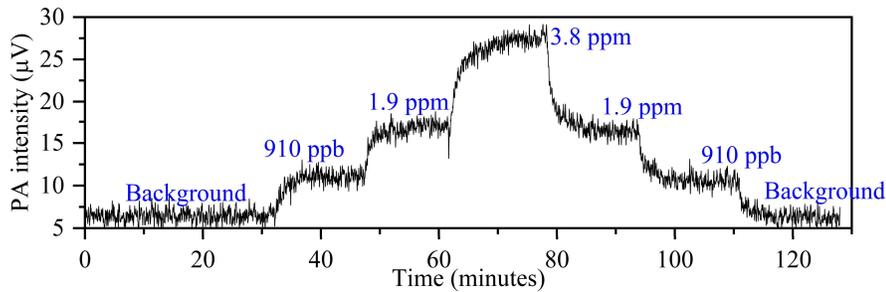

**Fig. S10 | Repeatability of PA detection of CO$_2$.**